\documentclass{ws-mpla}
\usepackage[super]{cite}
\usepackage{graphicx}
\begin{document}

\markboth{M.V. Bondarenco}
{Separation of Edge Effects in Highly Nondipole Radiation Spectra}

%%%%%%%%%%%%%%%%%%%%% Publisher's Area please ignore %%%%%%%%%%%%%%
\catchline{}{}{}{}{}
%%%%%%%%%%%%%%%%%%%%%%%%%%%%%%%%%%%%%%%%%%%%%%%%%%%%%%%%%%%%%%%%%%%

\title{SEPARATION OF EDGE EFFECTS IN HIGHLY NON-DIPOLE RADIATION SPECTRA}

\author{\footnotesize M.V. BONDARENCO
%\footnote{
%Typeset names in 8 pt Times Roman, uppercase. Use the footnote to
%indicate the present or permanent address of the author.}
}

\address{
NSC Kharkov Institute of Physics and Technology, 1 Academic St.\\
Kharkov, 61108, Ukraine\\
V.N. Karazine Kharkov National University, 4 Svobody Sq.\\
Kharkov, 61077, Ukraine
%\footnote{State completely without abbreviations, the
%affiliation and mailing address, including country and e-mail address.
%Typeset in 8 pt Times Italic.}
\\
bon@kipt.kharkov.ua}

%\author{SECOND AUTHOR}

%\address{Group, Laboratory, Address\\
%City, State ZIP/Zone, Country
%}

\maketitle

\pub{Received (Day Month Year)}{Revised (Day Month Year)}

\begin{abstract}
Edge effects in spectra of non-dipole radiation from ultra-relativistic electrons passing through various finite targets are analyzed from a unified point of view. Examples include radiation in a finite magnet, bremsstrahlung at double scattering, and bremsstrahlung in an amorphous plate. A generalization of the encountered spectral decomposition property is proposed. The relevance of phenomena of jet-interjet radiation interference and electron time delay in the target is emphasized. Features produced by them in the radiation spectra are discussed.

\end{abstract}

\ccode{PACS Nos.: 13.85.Qk, 12.38.Bx, 12.15.Ji}

\section{Introduction}

Spectra of radiation from ultra-relativistic electrons in finite targets often contain appreciable edge effects. One of the common known
examples  is transition radiation on discontinuities of dielectric susceptibility, and its interference between the boundaries of a traversed plate
\cite{Ginzburg,trans-rad-interf}. Less well-known but not less important are  effects related to actual breaks in the electron trajectory -- e.g., radiation from an electron in a gap between the magnets in
a storage ring \cite{synchr-rad-straight-section}, or for an electron circumscribing a finite arc in a bending magnet \cite{Bagrov-Fedosov-Ternov}.

Studies of edge effects in direct radiation historically begun with the simplest problems, when electron deflection angles are either small enough (dipole radiation), or of the order of unity. More recently, attention was drawn to situations when angles of deflection of a high-energy electron in the target are small compared to unity, but well exceed the inverse Lorentz factor serving as the scale for radiation emission angles. Under such conditions, the radiation is forward-peaked, but highly non-dipole. Conventionally, measured at that are radiation spectra integral over photon emission angles.

\begin{figure}
\includegraphics{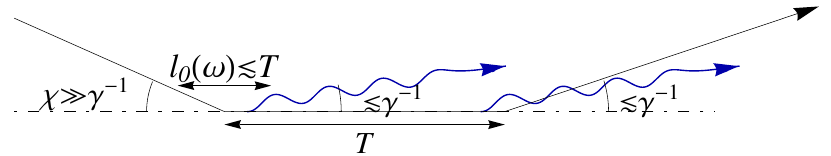}
 \caption{\label{fig:Diagr1} Diagram for collinear-collinear radiation interference from double hard scattered electron (high-$\omega$ spectral region).
%(a), and low-$\omega$ part of the spectrum (b).
%In the latter case there is also a cross-diagram, in which loose
%photons are emitted from the first scattering vertex, and collinear
%ones from the final electron line
}
\end{figure}

An impetus to studies of nondipole radiation from high-energy electrons was given by recent experimental investigation of radiation at double scattering \cite{NA63-plans}, aiming at verification of predictions \cite{Blankenbecler,Zakharov,BK-structured} about interference fringes in spectra of bremsstrahlung on two amorphous foils. Such fringes may basically be described by classical electrodynamics (granted that typical photon energies $\hbar\omega$ are much lower than the electron energy). Embarking at the textbook formula for spectral distribution of irradiated energy\begin{eqnarray}\label{dIdomega-through-angles}
\frac{dI}{d\omega}=\omega^2\int d^2n
\left|\frac{e}{2\pi}\int_{-\infty}^{\infty} dt
[\vec{n},\vec{v}(t)]e^{i\omega t-i\vec{k}\cdot\vec{r}(t)}\right|^2,
\end{eqnarray}
(we let $c=1$), one can bring it to a form of a single integral \cite{Bondarenco-Shulga}
\begin{equation}\label{dIdomega-combined}
\left\langle\frac{dI}{d\omega}\right\rangle_{1,2}=\left\langle\frac{dI_{\text{BH}}}{d\omega}\right\rangle_1+\left\langle\frac{dI_{\text{BH}}}{d\omega}\right\rangle_2%\qquad\qquad\qquad\qquad\nonumber\\
-\frac{2e^2\gamma^4}{\pi}\!\int_0^{\infty}\!d\theta^2\theta^2\left\langle
G\right\rangle_1\!\left\langle
G\right\rangle_2\cos\left[\frac{\omega T}{2\gamma^2}(1+\gamma^2\theta^2)\right],% \,\,
\end{equation}
where $\gamma=1/\sqrt{1-v^2}$ is the electron Lorentz factor, $\theta$ the radiation emission angle with respect to the intermediate electron line,
\begin{equation*}\label{J-scalar-def}
G(\theta)\approx \frac{1}{1+\gamma^{2}\theta^2}%\qquad\qquad\qquad\qquad\qquad\qquad\quad\qquad\nonumber\\
-\frac{1}{
2\gamma^{2}\theta^2}\left(1+\frac{\theta^2-\chi^2-\gamma^{-2}}{\sqrt{[\gamma^{-2}+(\theta-\chi)^2][\gamma^{-2}+(\theta+\chi)^2]}}\right),
\end{equation*}
and $\left\langle ...\right\rangle$ denotes the averaging over
electron scattering angles $\chi_1$ and $\chi_2$ in each foil. The
last term in (\ref{dIdomega-combined}) gives rise to oscillatory
spectral behavior
\begin{equation}\label{dI-largelambda}
\left\langle\frac{dI}{d\omega}\right\rangle_{1,2}\underset{\omega\gg\gamma^2/T}\simeq\left\langle\frac{dI_{\text{BH}}}{d\omega}\right\rangle_1+\left\langle\frac{dI_{\text{BH}}}{d\omega}\right\rangle_2%\qquad\qquad\quad\nonumber\\
+\frac{8e^2}{\pi}\left\langle G(0)\right\rangle_1\left\langle
G(0)\right\rangle_2 \left(\frac{2\gamma^2}{\omega T}\right)^2 \cos
\frac{\omega T}{2\gamma^2},
\end{equation}
\begin{equation}\label{I-BH}
\frac{dI_{\text{BH}}}{d\omega}(\gamma\chi)\underset{\gamma\chi\gg1}\simeq\frac{2e^2}{\pi}(\ln\gamma^2\chi^2-1)
\end{equation}
[with $G(0)=1-\frac1{\left(1+\gamma^2\chi^2\right)^{2}}$], which is similar to that for interference
of transition radiation from different boundaries of the traversed
plate \cite{trans-rad-interf}, or to that for radiation in a gap of a storage ring \cite{synchr-rad-straight-section}. The phase of the cosine in
Eq.~(\ref{dI-largelambda}) may be understood as the physical ratio
${\omega T}/{2\gamma^2}=T/l_0(\omega)$, where
\begin{equation}\label{l0-def}
l_0(\omega)=\frac{2\gamma^2}{\omega}
\end{equation}
is the ``free" photon formation length. Notably, oscillations
(\ref{dI-largelambda}) enhance in the highly non-dipole regime
$\gamma^{-1}\ll\chi\ll1$, when
$G\underset{\chi\gg\gamma^{-1}}\to\frac1{1+\gamma^2\theta^2}$,
becoming independent of the scattering angles which are averaged
over. That is not unnatural, as long as in the opposite, dipole
limit their average must vanish (dipole Bethe-Heitler contributions
do not interfere on the average).

%Interference of the described type between the endpoints of the internal
%rectilinear segment of electron's trajectory may be regarded as a manifestation of the so-called ``semi-bare electron" (see Fig.~\ref{fig:Diagr1}).
One may ask, however, why does the phase of the interference term
depend on $T/l_0(\omega)$ alone, but is independent of the electron scattering angles.
It might as well depend on the scattering-modified photon formation
length
\begin{equation}\label{lchi-def}
l_{\chi}(\omega)\simeq l_f(\omega,\theta)\big|_{\theta\sim\chi}\equiv\frac2{\omega(\gamma^{-2}+\theta^2)}\bigg|_{\theta\sim\chi}\sim\frac2{\omega\chi^2}\qquad\quad (\chi\gg\gamma^{-1})
\end{equation}
relevant, e.g., in the theory of Landau-Pomeranchuk-Migdal effect
\cite{Galitsky-Gurevich}, which, after all, is just another
manifestation of radiation interference. A closer examination
reveals that in the two-foil case, interference effects related with
length (\ref{lchi-def}) are erased by averaging. But for problems
without averaging, they may be viable and observable.

Whereas from Fig.~\ref{fig:Diagr1} it is clear which kinds of photons
are interfering on scale $l_0(\omega)$, for scale $l_{\chi}(\omega)$
that may be less obvious. In what follows, we shall demonstrate that the
interference at low $\omega$ involves nontrivial geometry
both in the longitudinal and transverse directions, and engages not one
but two categories of photons: intra-jet (inside a jet formed by a
temporarily straight moving electron) and inter-jet (between the
jets) \cite{Bond-Shul-double-scat}. Therefore, the mechanism of edge
radiation interference at low $\omega$ is not the same as at high $\omega$, but more intricate.

At the same time, all highly-nondipole radiation problems in finite targets have much in common. In this paper, we propose a generalization for their spectral decomposition. It seems expedient to begin with its statement.

\section{Separation of volume and edge contributions for nondipole radiation in finite targets}\label{sec:decomposition}

To assess edge effects, first of all, it is imperative to define how to
discriminate them from the volume contribution. A natural and generally applicable procedure is to indefinitely expand the target (i.e., formally send its size
$T\to\infty$ with the rest of the parameters held fixed), and split
\begin{equation}\label{vol+edge}
\frac{dI}{d\omega}=\frac{dI_{\text{vol}}}{d\omega}+\frac{dI_b}{d\omega},
\end{equation}
where $\frac{dI_{\text{vol}}}{d\omega}\propto T$ [i.e., $\frac{dI_{\text{vol}}}{d\omega}=T\left(\underset{T\to\infty}\lim\frac1T\frac{dI}{d\omega}\right)$], and $\frac{dI_b}{d\omega} \underset{T\to\infty}=O(1)$.

The leading, ``volume" contribution $\frac{dI_{\text{vol}}}{d\omega}$ is expected to be generated ``locally"
(i.e., at spatial scales much smaller than $T$). So, it must depend on
$\omega$ essentially through the ratio
\begin{equation}\label{hard-scale}
\frac{l_{\text{ext}}}{l_0(\omega)}.
\end{equation}
Here $l_{\text{ext}}=\max\tau(\chi\lesssim\gamma^{-1})$ is the time scale within which the electron deflection angles do not overwhelm typical radiation emission angles $\sim\gamma^{-1}$, and thereby the radiation coherence is maintained. Hence, $l_{\text{ext}}$ plays the role of the external field coherence length.

The remainder $\frac{dI_b}{d\omega}$, which must embody all the edge effects, should depend on variable
(\ref{hard-scale}) somewhat differently, because it is generated in a physically different way. Besides that, it can depend on a
ratio
\begin{equation}\label{soft-scale}
\frac{T}{l_{\chi}(\omega)}.
\end{equation}
But when scales of $\omega$, at which variables (\ref{hard-scale}) and (\ref{soft-scale}) are of the order of unity,
vastly differ ($\chi\gg\gamma^{-1}$), the dependence
of $\frac{dI_b}{d\omega}$ on them may actually be treated as additive.

To gain more insight into the structure of edge effects, it should be realized that in the angular distribution of highly nondipole radiation on a finite target there are always prominent jets. They are related with formation of photons attached to the initial or the final electron line during long times, and thereby being narrowly collimated (to within angles $\sim\gamma^{-1}$ around the electron velocity).\footnote{The term ``jet" is widely used in high-energy hadron physics, where it signifies production of many hadrons of different species and energies in about the same direction. Since the advent of QCD, though, such a term is also applied on the quantum field theoretical, parton level, often involving just a quark and a collinear gluon emitted by it. A similar nomenclature could be extended as well to electrodynamics, where a ``jet" consists of an electron and a photon collinear to it (see also \cite{Carimalo-Schiller-Serbo}).} Besides that, radiation persists also at angles between the jets, being fainter but filling a wider angular region (see, e.g., Fig. 2 of \cite{Bond-Shul-double-scat}). Even though in radiation spectra the emission angles are integrated over, jet effects can survive under conditions of nondipole radiation. In a generic case considered herein, these jets do not overlap in the plane of emission angles, and thus do not interfere, but jet photons from one electron line can interfere with interjet photons emitted at the opposite boundary of the target.

Under the given conditions, the formal additivity of manifestations of the two scales in the radiation spectrum has a clear-cut physical meaning. In decomposition
%The prime reason is that even though intra-jet and inter-jet photons can interfere (see Fig.~\ref{fig:Diagr} below), the photon formation length in that case is the shortest among them, anyway. Therefore,
\begin{equation}\label{hard+soft}
\frac{dI_b}{d\omega}\underset{\chi\gg\gamma^{-1}}\simeq 2\frac{dI_{1b}}{d\omega}\left(\frac{l_{\text{ext}}}{l_0(\omega)}\right)+\frac{dI_{bb}}{d\omega}\left(\frac{T}{l_{\chi}(\omega)}\right),\qquad \frac{dI_{bb}}{d\omega}\underset{T/l_{\chi}\to0}\to0,
\end{equation}
$\frac{dI_{1b}}{d\omega}$ corresponds to a single-boundary contribution, and is normally independent of $T$, as well as of the wide
deflection angles collectively denoted here as $\chi$. On the other hand,
$\frac{dI_{bb}}{d\omega}$ corresponds to effects of interference between the boundaries, and appears to be basically independent of $\gamma$. More precisely, since $\frac{dI_{bb}}{d\omega}$ generally involves some jet effects, too, sp it bears a
mild $\gamma$-dependence, generally being expressible as\footnote{When the target boundaries are not identical, factor 2 at $A_1$ in Eq.~(\ref{dIsoft=2A1Fj+A2}), as well as factor 2 at $\frac{dI_{1b}}{d\omega}$ in Eq.~(\ref{hard+soft}), must be replaced by summation over the boundaries -- cf. Eq.~(\ref{dIsoftdomega-sum-formfactors}) below.}
\begin{equation}\label{dIsoft=2A1Fj+A2}
\frac{dI_{bb}}{d\omega}=\frac{2e^2}{\pi}\left[2A_1\left(\frac{\omega T\chi^2}{2} \right)F_{\perp}\left(\frac{\omega T\chi}{\gamma} \right)+A_2\left(\frac{\omega T\chi^2}{2} \right)\right].
\end{equation}
Functions $A_1$ and $A_2$ usually are oscillatory and decreasing as power laws. They may be called ``quasi-antenna" formfactors, where the role of the ``antenna" is played by the \emph{sufficiently bent} electron's trajectory (abstracting from the electron as a pointlike particle), which may be treated as a long ``wire", along which the electric current evoked by the electron motion flows essentially at the speed of light
($\gamma\to\infty$). In turn, $F_{\perp}$ is a monotonously decreasing proper field formfactor, absorbing all the jet effects, and furnishing exponential decrease of the oscillations at intermediate $\omega$.

In separation (\ref{hard+soft}), single and double boundary
contributions are logarithmically divergent individually (cf., e.g., \cite{Goldman}),
\begin{equation}\label{pm-Log-omega}
2\frac{dI_{1b}}{d\omega},
\frac{dI_{bb}}{d\omega}\underset{\omega\to0}\sim\pm\ln\frac1{\omega},
\end{equation}
but the divergences cancel in their
sum.

The quoted formulas pertain to the case when the target edges are
sharp. To take their non-zero width $\Delta T\ll T$ into account, one has
to replace
\begin{equation}\label{}
\frac{dI_{1b}}{d\omega}\to \frac{dI_{1b}}{d\omega} F_{\text{edge}}(\omega\Delta T/\gamma^2),
\end{equation}
with an additional formfactor $F_{\text{edge}}$, whereas
contributions $\frac{dI_{\text{vol}}}{d\omega}$,
$\frac{dI_{bb}}{d\omega}$ are unaltered.

\section{Non-dipole spectral decompositions for specific problems}

\subsection{Radiation in a finite magnet}\label{subsec:magnet}

\begin{figure}
    \includegraphics{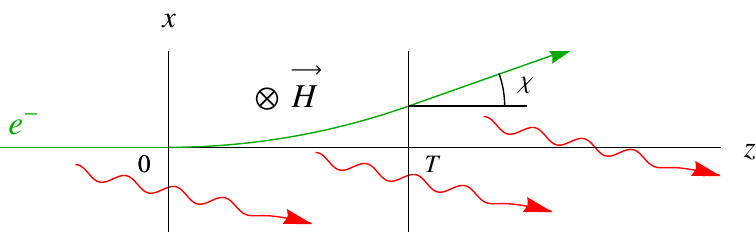}
\caption{\label{fig:magnet} Geometry of radiation from an electron
in a finite magnet. Photon emissions from the three distinct
spatial regions are generally interfering.}
\end{figure}

To illustrate the conjectured decomposition, consider
a few specific examples. First, consider a process of radiation from a fast electron passing through a long but finite magnet
 (see Fig.~\ref{fig:magnet}). This physical problem was studied on general grounds in \cite{Bagrov-Fedosov-Ternov},
but our objective is to decouple single-edge and edge interference contributions based on the long-magnet asymptotics.
%Such a problem was discussed in \cite{Coisson,Bagrov-F

In this case, for all $\omega$ it is advantageous to employ
for the radiation spectrum the double time integral representation
[see \cite{BKS} and Eq.~(\ref{dIdomega-photon-proparator}) below].
The result of the integrations gives structure (\ref{vol+edge}), (\ref{hard+soft}) with entries
\begin{equation}\label{dI-synch}
\frac{dI_{\text{vol}}}{d\omega}=2e^2X\left\{-(2\Omega_s)^{1/3}\text{Ai}'\left[(2\Omega_s)^{2/3}\right]%\nonumber\\
-\Omega_s\int_{(2\Omega_s)^{2/3}}^{\infty}d\alpha
\text{Ai}(\alpha)\right\}
\end{equation}
(representing the synchrotron radiation intensity times the magnet
length,  $\Omega_s=\frac{\omega
R}{2\gamma^3}$, $X=\frac{\gamma T}{R}=\gamma\chi$, $R$ stands for the trajectory bending radius),
\begin{eqnarray}\label{Iinterf-Gi}
\frac{\pi}{e^2}\frac{dI_{1b}}{d\omega}=(2\Omega_s)^{\frac23}\pi\text{Gi}\left[(2\Omega_s)^{\frac23}\right]-3\qquad\qquad\qquad\qquad\qquad\qquad\qquad\qquad\qquad\qquad\nonumber\\
+\int_1^{\infty}\!\!\!\frac{dw}{w-\frac34}\Bigg\{1%\qquad\qquad\qquad\qquad\qquad\qquad\qquad\nonumber\\
+\Omega_s^{\frac23}\!\left[2\!\left(\!1-\frac3{4w}\right)^{\!\frac23}
\!\!-w\left(\!1-\frac3{4w}\right)^{\!-\frac13}\right]%\nonumber\\ \times
\!\pi\text{Gi}\!\left(\Omega_s^{\frac23}w\!\left(\!1-\frac3{4w}\right)^{\!-\frac13}\right)\!\!\Bigg\},\nonumber\\
\qquad\qquad
\end{eqnarray}
where $\text{Gi}(\alpha)=\frac1{\pi}\int_0^{\infty}dt \sin\left(\alpha t+\frac13 t^3\right)$ is the Scorer function, arising naturally instead of the Airy function $\text{Ai}(\alpha)=\frac1{\pi}\int_0^{\infty}dt \cos\left(\alpha t+\frac13 t^3\right)$ in description of single-edge effects, and finally,
\begin{equation}\label{boundary-interf-part-magnet}
\frac{dI_{bb}}{d\omega}=\frac{2e^2}{\pi}\left[2A_{1}\left(\Omega_s
X^3\right)F_{\perp}(\Omega_s X^2)+A_{2}\left(\Omega_s X^3\right)\right],
\end{equation}
where
\begin{equation}\label{Fj-def}
F_{\perp}(z)=zK_1(z),\qquad F_{\perp}(0)=1
\end{equation}
is a proper field form factor \cite{Bond-Shul-double-scat}, whose relevance for the present
process will be elucidated in Sec~\ref{subsec:imp-par},
whereas antenna formfactors are
\begin{equation}\label{A1-magn}
A_{1}=\frac{2}{\Omega_s
X^3}\int_0^{\infty}\frac{du}{(1+u)^2}%\qquad\qquad\qquad\qquad\quad\nonumber\\ \times
\Bigg\{\sin\left[\frac{\Omega_s X^3}{3}(1+u)\right]
-\sin\left[\frac{\Omega_s
X^3}{2}\left(\frac23+u\right)\right]\Bigg\},
\end{equation}
\begin{eqnarray}\label{A2-magn}
A_{2}=\int_0^{\infty}\frac{du}{1+u}\Bigg\{\cos\left[\frac{\Omega_s X^3}{12}(1+3u)\right]-\cos\left[\frac{\Omega_s X^3}{12}(1+u)\right]\nonumber\\
+\frac2{1+u}\cos\left[\frac{\Omega_s X^3}{12}(1+u)^3\right]\Bigg\}.
\end{eqnarray}

In spite of some bulkiness of the encountered formulae, the present case is the most ``normal" manifestation of decomposition (\ref{vol+edge}), (\ref{hard+soft}): Structure (\ref{boundary-interf-part-magnet}) exactly corresponds to conjectured representation (\ref{dIsoft=2A1Fj+A2}), while (\ref{Iinterf-Gi}) depends solely on $\Omega_s$, which may be cast in form (\ref{hard-scale}), (\ref{l0-def}) with $l_{\text{ext}}=R/\gamma$.

Single-boundary contribution $\frac{dI_{1b}}{d\omega}$ (an analog of transition radiation, arising in the absence of atomic matter) is not positive definite, so, it does not give an independent radiation intensity, and needs a sufficient volume radiation background.
Its salient feature is that at high $\omega$, the edge contribution falls off as a (positive) power law
\begin{equation}\label{highomega}
\frac{dI}{d\omega}\underset{\Omega_s\to\infty}\simeq 2\frac{dI_{1b}}{d\omega}\simeq \frac{7e^2}{15\pi\Omega_s^2},
\end{equation}
which must ultimately dominate over the exponentially attenuating volume synchrotron contribution. Qualitative prediction of asymptotic behavior
$\frac{dI}{d\omega}\underset{\omega\to\infty}\sim\omega^{-2}$
was made in \cite{Bagrov-Fedosov-Ternov}.%, while Eq.~(\ref{highomega}) quantifies its coefficient.

At low $\omega$, terms (\ref{Iinterf-Gi}) and (\ref{A1-magn}) combine to mutually cancel their logarithmic singularities [cf. Eq.~(\ref{pm-Log-omega})] and ensure the correspondence with the factorization theorem.
%synchrotron contribution (\ref{dI-synch}) vanishes, whereas the edge contribution does not.
%Edge contributions therefore dominate both at high and low $\omega$.

A related problem arises for radiation at electron passage through a short crystal (see, e.g., \cite{edge-crystal}), particularly when the electron passing through
the inter-planar channel nearly conserves its impact parameter,
thus experiencing a constant transverse force. In that case, the
target edges are sharper than those for laboratory magnets, but
there arises an additional need for averaging over electron impact
parameters.

%\begin{figure}
%    \includegraphics[width=67mm]{Low-Omega}
%\caption{\label{fig:Low-Omega} Log plot of the radiation spectrum in the low-$\Omega_s$ region, for $X=30$ (solid curve). Dashed curve, pure $X J_{\text{syn}}$.}
%\end{figure}

\subsection{Bremsstrahlung at double hard scattering}\label{subsec:double-scattering}

%\begin{figure}
%\includegraphics{traj-scheme}
% \caption{\label{fig:traj-scheme} Geometry of electron double scattering and the accompanying radiation. }
%\end{figure}

\begin{figure}
\includegraphics{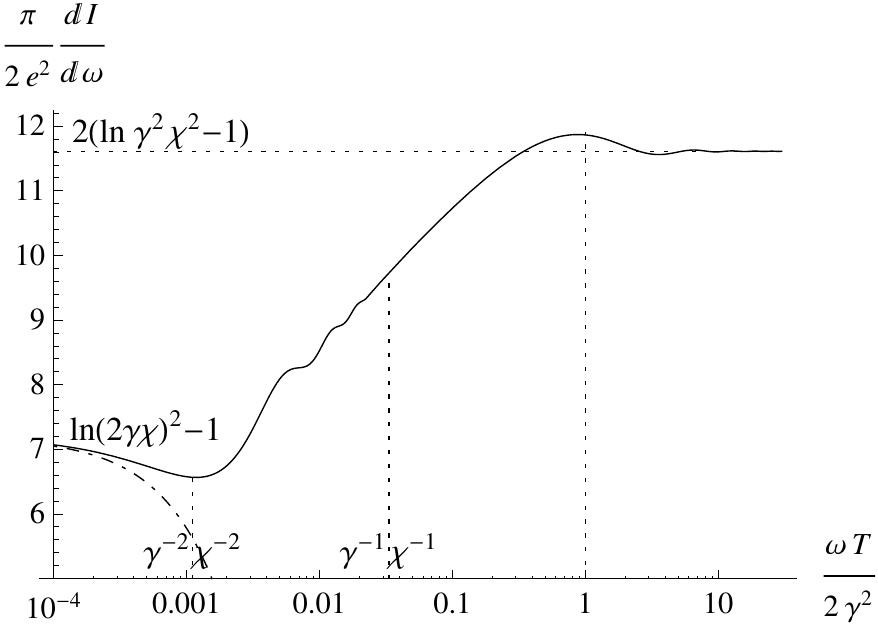}
\caption{\label{fig:Oscill} (Adapted from Ref.~\cite{Bond-Shul-double-scat}.) Spectrum of electromagnetic radiation from an electron scattering two times through co-planar ($\varphi_{12}=0$) equal angles $\chi=30\gamma^{-1}$. Dot-dashed curve, Eq.~(\ref{dIdomega-O(omega)}).}
\end{figure}

A subtler example, already mentioned in the Introduction, arises when instead of interaction with a continuous target, the electron undergoes two instantaneous scatterings. Specifically, let the electron scatter repeatedly, at points separated by a time interval $T$, through angles
$\vec{\chi}_1$, $\vec{\chi}_2$ with a relative azimuth
$\varphi_{12}=\arccos
\left(\vec{\chi}_1\cdot\vec{\chi}_2/|\vec{\chi}_1||\vec{\chi}_2|\right)$,
such that $|\vec{\chi}_1|,|\vec{\chi}_2|\gg\gamma^{-1}$. After
appropriate integrations in (\ref{dIdomega-through-angles}), one
arrives \cite{Bond-Shul-double-scat} at the expression for the radiation spectrum (\ref{vol+edge}), (\ref{hard+soft}), with
\begin{equation}\label{Ivol-IBH}
\frac{dI_{\text{vol}}}{d\omega}=\sum_{i=1}^{2}\frac{dI_{\text{BH}}}{d\omega}(\gamma\chi_i),
\end{equation}
\begin{equation}\label{Ihard-2scat}
\frac{dI_{1b}}{d\omega}=-\frac{e^2}{\pi}\int_0^{\infty}
\frac{d\theta^2\theta^2}{(\gamma^{-2}+\theta^2)^2}\cos\left[\frac{\omega
T}{2\gamma^2}\left(1+\gamma^2\theta^2\right)\right]
\end{equation}
[the latter being similar to the non-dipole limit of the \emph{non-averaged} interference term in (\ref{dIdomega-combined})], and
\begin{eqnarray}\label{dIsoftdomega-sum-formfactors}
\frac{dI_{bb}}{d\omega}=
\frac{2e^2}{\pi}\Bigg[A_1\left(\frac{\omega T\chi_1^2}2,\frac{\omega T\chi_1{\chi}_2}2 e^{i\varphi_{12}}\right)F_{\perp}\left(\frac{\omega T\chi_1}{\gamma}\right)\qquad\qquad\qquad\qquad\nonumber\\
+A_1\left(\frac{\omega T\chi_2^2}2,\frac{\omega T\chi_1{\chi}_2}2
e^{i\varphi_{12}}\right)F_{\perp}\left(\frac{\omega
T\chi_2}{\gamma}\right)%\nonumber\\
+A_2\left(\frac{\omega T\chi_1{\chi}_2}2
e^{i\varphi_{12}}\right)\Bigg]\quad %\nonumber\\
\end{eqnarray}
with
\begin{equation}\label{A1-def}
A_1\left(z_1,z_2\right)=-\text{Ci}\left(z_1\right)%\qquad\qquad\qquad\qquad\qquad\quad\nonumber\\
+\mathfrak{Re}\left\{\cos z_2\text{Ci}\left(z_1+z_2\right)+\sin
z_2\text{si}\left(z_1+z_2\right)\right\},
\end{equation}
\begin{equation}\label{A2-def}
A_2\left(z\right)=-\mathfrak{Re}\left\{\cos z
\text{Ci}\left(z\right)+\sin z\text{si}\left(z\right)\right\},
\end{equation}
$\text{Ci}(z)=-\int_z^{\infty}\frac{dx}{x}\cos x$,
$\text{si}(z)=-\int_z^{\infty}\frac{dx}{x}\sin x$, and $F_{\perp}$ given by Eq. (\ref{Fj-def}). Contributions (\ref{Ihard-2scat}) and (\ref{A1-def}), (\ref{A2-def}), depending on different photon formation
lengths, exhibit oscillations in different spectral regions (see
Fig.~\ref{fig:Oscill}). Presently, only hard oscillations (\ref{Ihard-2scat}), corresponding to formation length $l_0$ were investigated experimentally \cite{NA63-plans}.

%What is the physical difference between $A_1$ and $A_2$?

Comparison of the obtained structures with generic
Eqs.~(\ref{hard+soft}), (\ref{dIsoft=2A1Fj+A2}) indicates that
(\ref{dIsoftdomega-sum-formfactors}) is perfectly consistent with
(\ref{dIsoft=2A1Fj+A2}). However, since now $l_{\text{ext}}=\max\tau(\chi\lesssim\gamma^{-1})=T$ (in spite that angles $\sim\chi$ are also achieved at scattering in single points), (\ref{Ihard-2scat}) depends on $T$,
and therefore it is not quite a single-boundary contribution.
Still, it can be treated as such, in the sense that the photons are formed only on one of the boundaries, but contributions from different boundaries can interfere. But a $T$-dependent scale can already not be relevant for $\frac{dI_{\text{BH}}}{d\omega}$. Fortunately, since the latter is actually $\omega$-independent [see Eq. (\ref{I-BH})], the problem does not exist. Physically, the encountered anomaly owes to the straightness of the electron motion in between the scattering points, so that even if photons are formed at its ends pretty far from each other, they move in the same direction and hence can interfere, as was depicted in Fig. \ref{fig:Diagr1}.

A more serious issue arises when initial and final electrons are
collinear [see Fig. 5(c) of \cite{Bond-Shul-double-scat}]. Such effects are not taken into account by the decomposition of Sec. \ref{sec:decomposition}. In particular, they should include also an overlap of two formfactors $F_{\perp}$, corresponding to initial and final electron states. But the latter case may be regarded as exceptional.

\section{Space-time analysis}

Proper understanding of the structure of the encountered spectral
decomposition (\ref{dIsoft=2A1Fj+A2}) requires studying its
space-time origin. Below we will explain in a nutshell the physical
mechanisms behind this structure.

\subsection{Intermediate $\omega$. Ray optics}\label{subsec:imp-par}

As was already mentioned in the Introduction, at high $\omega$ the interference is governed by the longitudinal coherence with length $l_0$.
With the decrease of $\omega$, transverse wavelengths increase and become comparable with the geometrical transverse scales. So, there increases the importance of proper treatment of transverse spatial dimensions.
They can be brought out, e.g., by passing, via Fourier transformation, from radiation emission angles to photon impact (or emission) parameters. For single scattering, such a photon impact parameter representation is rather well-known \cite{Bjorken-Kogut-Soper,Zakharov,BK-LPM}:
%\begin{subequations}
\begin{equation}\label{K1^2(1-J)}
\frac{dI}{d\omega } =\left(\frac{e}{\pi}\right)^2\int d^2\xi
\left[\frac{\partial}{\partial
\vec{\xi}}K_0(\xi/\gamma)\right]^2\left|1-e^{i\vec{\chi}\cdot
\vec{\xi}}
\right|^2=\frac{dI_{\text{BH}}}{d\omega}(\gamma\chi),%\nonumber\\
%&\,&\label{40a}\\
%&=&\frac{4e^2}{\pi\gamma^2}\int_0^{\infty} dr r
%K_1^2(r/\gamma)\left[1-J_0\left(\chi
%r\right)\right].
\end{equation}
%\end{subequations}
where $\vec{b}=\vec{\xi}/\omega$ is the impact parameter.
Integration in (\ref{K1^2(1-J)}) recovers form (\ref{I-BH}).

\begin{figure}
\includegraphics{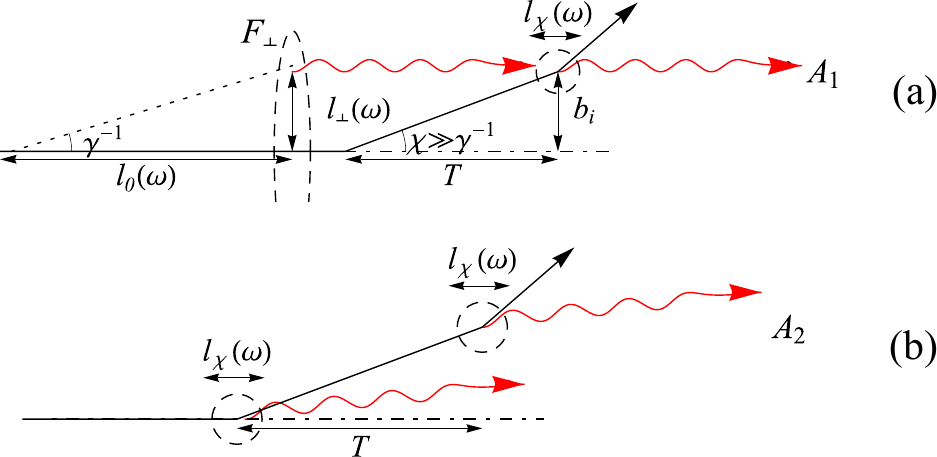}
 \caption{\label{fig:Diagr} (a) Graphical illustration of formation of the jet-interjet interference contribution $A_1$ and its modulating formfactor $F_{\perp}$.
The condition of interference
between collinear and noncollinear photons, besides the
coincidence of emission directions, is the equality of impact
parameters: $l_{\perp}(\omega)={\gamma}/{\omega}$ and $b_i\sim T\chi$. There is also a cross diagram, in which interjet
photons are emitted from the first scattering vertex, and jet
ones from the final electron line.
(b) The same for interjet-interjet interference contribution $A_2$.}
\end{figure}

An extension of this approach to double scattering leads to representation
\cite{Bondarenco-Shulga,Bond-Shul-double-scat}
\begin{eqnarray}\label{imp-par}
\frac{dI}{d\omega}=\frac{dI_{\text{BH}}}{d\omega}(\gamma\chi_1)+\frac{dI_{\text{BH}}}{d\omega}(\gamma\chi_2)\qquad\qquad\qquad\qquad\qquad\qquad\nonumber\\
-\frac{e^2}{\pi^3\omega T}\iint d^2 \xi_1 d^2 \xi_2
\frac{\partial}{\partial
\vec{\xi}_1}K_0\left(\frac{\xi_1}{\gamma}\right)
\cdot\frac{\partial}{\partial
\vec{\xi}_2}K_0\left(\frac{\xi_2}{\gamma}\right)\qquad
 \nonumber\\
\times \mathfrak{Im}\left\{\left(1-e^{-i\vec{\chi}_1\cdot \vec{\xi}_1}
\right)\left(1-e^{-i\vec{\chi}_2\cdot \vec{\xi}_2} \right)
e^{-i\frac{\omega
T}{2\gamma^{2}}+i\frac{(\vec{\xi}_1-\vec{\xi}_2)^2}{2\omega T}}\right\}.
\end{eqnarray}
On its basis, it is straightforward to demonstrate the onset of ray
optics in the given process: When $\chi\gg\gamma^{-1}$,
exponentials $e^{-i\vec{\chi}_1\cdot \vec{\xi}_1}$,
$e^{-i\vec{\chi}_2\cdot \vec{\xi}_2}$ are rapidly oscillating, and
along with the Gaussian factor
$e^{i\frac{(\vec{\xi}_1-\vec{\xi}_2)^2}{2\omega T}}$, they form
(real) stationary phase points, defining rays parallel to one of the
external electron lines. The monotonous decrease of the
impact-parameter-dependent photon distributions at these impact
parameters gives rise to the proper field formfactor for $A_1$ -- see
Fig.~\ref{fig:Diagr}(a).\footnote{For transition radiation such
effects are impossible in principle, because there the charged
particle trajectory is everywhere rectilinear.} On the contrary,
Fig.~\ref{fig:Diagr}(b) demonstrates the absence of a modulating formfactor for $A_2$.

%\begin{figure}
%    \includegraphics{critical-imp-par}
%\caption{\label{fig:critical-imp-par} Retarded electromagnetic field of a relativistic electron orbiting in a synchrotron with $v=0.9c$. The strip of the radiation field occurs at a finite transverse distance wrt electron motion.}
%\end{figure}

In the case of radiation in a magnet, it should be remembered yet that synchrotron photons within the magnet are stripped at a non-zero intrinsic impact parameter\footnote{Transverse scale (\ref{Delta-x-syn}) is related via $\Delta x_{\text{syn}}=t_{\text{syn}}^2/2R$ with the longitudinal scale such that $t_{\text{syn}}\sim l_{\chi}[\chi^2(t_{\text{syn}})]$, where $\chi(t)=t/R$ and $l_{\chi}(\chi)=1/\omega\chi^2$, as given by Eq.~(\ref{lchi-def}). Another characteristic transverse parameter \cite{Artru-imp-par-synchr}, equal to $R/2\gamma^2$, under the present conditions is smaller.}
\begin{equation}\label{Delta-x-syn}
\Delta x_{\text{syn}}=\xi_x/\omega\sim R^{1/3}\omega^{-2/3}.
\end{equation}
At small $\omega$, that quantity is large, but with the increase of $\omega$, it ultimately falls below the impact parameter difference between electron entrance and exit from the magnet:
\begin{equation}
b_i=\left|\int_0^T dt\left[\vec{\chi}(t)-\vec{\chi}_i\right]\right|=T^2/2R,
\end{equation}
or
\begin{equation}
b_f=\left|\int_0^T dt\left[\vec{\chi}_f-\vec{\chi}(t)\right]\right|=T^2/2R,
\end{equation}
which are independent of $\omega$. That happens at $\Delta x_{\text{syn}}\ll b_i, b_f$, i.e., $\omega T\chi^2\gg1$, which is exactly the region where oscillations of $A_1$ develop. Then, $b_i, b_f$ become sufficiently sharply defined, so the electron proper field amplitude (\ref{Fj-def}) factors out with $b\to b_i, b_f$, and its exponential falloff at large $\omega$ eventually suppresses the spectral oscillations.

%Hence, at intermediate $\omega$ the impact parameter
%relationships should be basically the same as for double scattering
%considered in the previous subsection.
%There are still two photon jets pointing along initial and final electron lines. They are linked by a ..., which represents the volume contribution (see Fig.~\ref{fig:ang-magn}).

\subsection{Low $\omega$. ``Radio" contribution}

At sufficiently low $\omega$, the ray optics concepts break down, isolated points no longer play a distinguished role, and the entire electron trajectory radiates as a whole. At that, the foreground is taken by time aspects.

\begin{figure}
\includegraphics{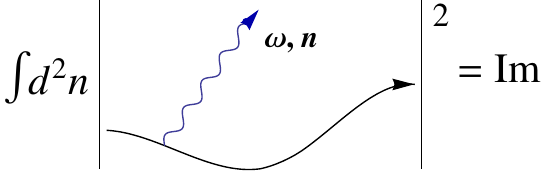}\includegraphics{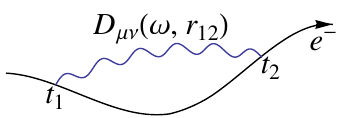}
 \caption{\label{fig:loop-diagram} Graphical illustration of unitarity relation (\ref{dIdomega-photon-proparator}).}
\end{figure}

Formally, at $\omega\to0$, there emerges a hierarchy of time scales, some of which, being reciprocal to $\omega$, expand indefinitely, whereas
others, determined by the target thickness, remain finite. To carry
out the corresponding scale separation self-consistently, it is convenient to embark at the representation for radiation spectrum, in which integration over photon emission angles is performed exactly. The resulting double
time integral representation expresses covariantly as
\cite{Bond-Shul-double-scat}
\begin{equation}\label{dIdomega-photon-proparator}
\frac1{\omega}\frac{dI}{d\omega}
= \frac{e^2}{\pi}\mathfrak{Im}  \int_{-\infty}^{\infty}ds_2\int^{s_2}_{-\infty}ds_1 u_{\mu}(t_1)u_{\nu}(t_2)%\qquad\nonumber\\ &\,&\times
e^{-i\omega
(t_2-t_1)}D_{\mu\nu}\left(\omega,\left|\vec{r}(t_2)-\vec{r}(t_1)\right|\right)
\end{equation}
and may be viewed as a kind of unitarity relation (see
Fig.~\ref{fig:loop-diagram}). Here $s=t/\gamma$ is the electron
proper time, $u_{\mu}$ its 4-velocity, and $D_{\mu\nu}$ the photon
propagator. Specializing $D_{\mu\nu}$ in the frequency-position
representation and Feynman gauge,
$D_{\mu\nu}(\omega,r)=-\frac{g_{\mu\nu}}{r-i0} e^{i\omega r}$,
reproduces the widely used formula \cite{BKS},
%\begin{eqnarray}\label{dIdomega-t2-t1+iepsilon}
%\frac{dI}{d\omega}= -\omega\frac{e^2}{\pi}\int_{-\infty}^{\infty}dt_2\int^{t_2}_{-\infty}dt_1 \left\{\gamma^{-2}+\frac12\left[\vec{v}(t_2)- \vec{v}(t_1)\right]^2\right\}\nonumber\\
%\times \mathfrak{Im}\frac{1}{t_2-t_1-i0} e^{-i\omega
%\left[t_2-t_1-\left|\vec{r}(t_2)-\vec{r}(t_1)\right|\right]},\quad
%\end{eqnarray}
whereas isolating the imaginary part of $\frac1{r-i0}$ and
presenting it in an integral form through a ``vacuum" term, to
representation \cite{Blankenbecler-Drell}.
%\begin{eqnarray}\label{t2tau-subtr}
%\frac{dI}{d\omega}
%=\omega\frac{e^2}{\pi}\int_0^{\infty}\frac{d\tau}{\tau}\int_{-\infty}^{\infty}\!dt_2\Bigg(\!\! \left\{\gamma^{-2}\!+\!\frac12\!\left[\vec{v}(t_2)- \vec{v}(t_2-\tau)\right]^2\right\}\nonumber\\
%\times \sin \omega \left[\tau-\left|\vec{r}(t_2)-\vec{r}(t_2-\tau)\right|\right]-\gamma^{-2} \sin\omega
%(1-v)\tau\Bigg).
%\end{eqnarray}

With the aid of formula (\ref{dIdomega-photon-proparator}), one can derive a next-to-leading order (NLO) correction to the low-$\omega$
approximation:
\begin{equation}\label{dIdomega-O(omega)}
\frac{dI}{d\omega}\underset{\omega\to0}\simeq
\frac{dI_{\text{BH}}}{d\omega}(\gamma\chi)+C_1\omega+\mathcal{O}(\omega^2),
\end{equation}
where the first (LO) term is the well-known factorization limit
given by Eq.  (\ref{I-BH}), and the coefficient at the NLO term
reads \cite{Bond-NLO}
\begin{equation}\label{C1}
C_1=-\frac{e^2}2 \int_{-\infty}^{\infty} dt
[\vec{\chi}(t)-\vec{\chi}_i]\cdot [\vec{\chi}_f-\vec{\chi}(t)],
\end{equation}
with $\vec{\chi}_i$ and $\vec{\chi}_f$ being the initial and final
electron deflection angles. The physical meaning of $C_1$ is that apart from the $e^2$ factor, it represents the difference
between the time delay for the actual trajectory and for its
angle-shaped approximation \cite{Bond-NLO}.

%At derivation of Eq. (\ref{C1}) it was used that for $C_1$ only one
%of the contributing times in (\ref{dIdomega-photon-proparator}) is
%large; this is why the NLO expansion for $dI/d\omega$ begins with
%$\mathcal{O}(\omega)$ [rather than $\mathcal{O}(\omega^2)$ as is the
%case for $dI/d\omega d^2n=|\mathcal{O}(1)+i\mathcal{O}(\omega)|^2$].

From Eq. (\ref{C1}), it is evident that for monotonous electron deflection (in particular, for cases considered in Secs.~\ref{subsec:magnet}, \ref{subsec:double-scattering}), always $C_1< 0$. Therefore, in those cases the spectrum suppression at low $\omega$ is non-monotonous.
%For passage through an amorphous target, one expects that $C_1=0$.

The salient feature of $C_1$ is its independence of $\gamma$ (or electron mass) for a definite trajectory $\vec{\chi}(t)$, i.e., definite particle energy and the field strength. Continuing the analysis under this assumption to all higher orders in $\omega$, one would recover complete antenna formfactors (\ref{dIsoft=2A1Fj+A2}), which are functions of a
$\gamma$- (or mass-) independent product $\omega T\chi^2$.

\section{Bremsstrahlung in an amorphous plate}\label{sec:amorph-plate}

Yet another, and perhaps practically most important example of a target with sharp boundaries is a solid plate (in the simplest case -- amorphous). The volume contribution in this case is described by
\begin{equation}\label{dIvol-amorph}
\frac{dI_{\text{vol}}}{d\omega}=
\left\langle \frac{dI_{\text{BH}}}{d\omega}\right\rangle \Phi_{M}(s),
\end{equation}
with $\left\langle \frac{dI_{\text{BH}}}{d\omega}\right\rangle=\frac{2e^2}{3\pi}\gamma^2\left\langle\chi^2\right\rangle$ and the Migdal function \cite{Migdal}
\begin{equation}\label{Phi-def}
\Phi_{M}(s)=6s^2\left\{4\mathfrak{Im}\psi\left[(1+i)s\right]-\frac1s-\pi\right\},\qquad \Phi_{M}(s)\underset{s\to\infty}\to1,
\end{equation}
involving $\psi(z)=\Gamma'(z)/\Gamma(z)$ and the argument related to ratio (\ref{hard-scale}) by
\begin{equation}\label{s-def}
8s^2=\frac{l_{\text{ext}}}{l_0(\omega)}=\frac{\omega}{2\gamma^4 d\left\langle\chi^2\right\rangle/d\tau}.
\end{equation}
The single-boundary contribution, evaluated in \cite{Goldman}, may be expressed as
\begin{equation}\label{1b-amorph-def}
\frac{dI_{1b}}{d\omega}=\frac{e^2}{\pi}B(2(1+i)s),
\end{equation}
\begin{eqnarray}\label{B-Goldman-def}
B(\sigma )&=&\mathfrak{Re}\Bigg\{2\sigma \int_0^{\infty}dx E_1\left[\sigma \tanh x\right] e^{-\sigma (x-\tanh x)}-2\nonumber\\
&\,&+\int_0^{\infty}dx e^{-\sigma  x}\left(1-\sigma  x\right)\left(\coth x-\frac1x\right)\Bigg\},
\qquad B(\sigma )\underset{s\to\infty}\to0,
\end{eqnarray}
and $E_1(z)=\int_{z}^{\infty}\frac{dx}{x}e^{-x}$.
It features a logarithmic infrared divergence. Boundary-boundary interference contribution can be read off from Eq.~(6.13) of \cite{BK-LPM}:
\begin{equation}\label{bb-amorph-def}
\frac{dI_{bb}}{d\omega}=\frac{2e^2}{\pi}A\left(\frac{\omega T\left\langle\chi^2\right\rangle}{2}\right),
\end{equation}
\begin{equation}\label{A-amorph-def}
A(z)=\ln\left(2|\sinh\sqrt{iz}|\right)-\sqrt{z/2},
\end{equation}
where the last term is the subtraction corresponding to the infrared square root singularity contained in (\ref{dIvol-amorph}), (\ref{Phi-def}), and also ensuring that $A(z)\underset{z\to\infty}\to0$.

\begin{figure}
\includegraphics{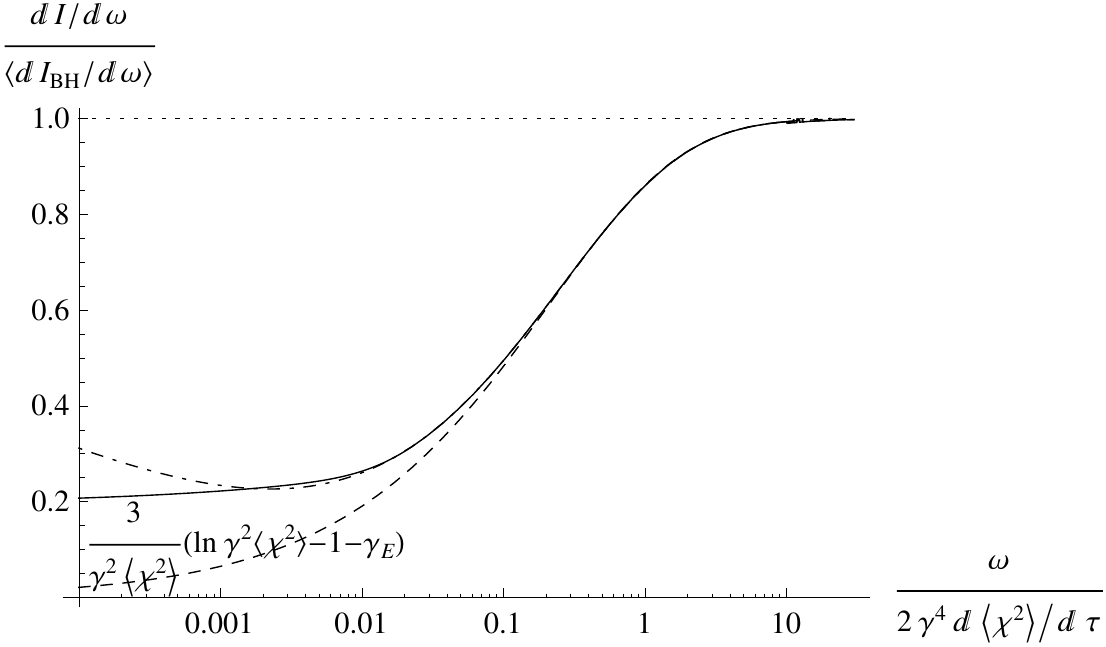}
 \caption{\label{fig:Amorph-spectrum} Normalized spectrum of bremsstrahlung on an amorphous plate, in which the final rms deflection angle equals $\sqrt{\left\langle\chi^2\right\rangle}=5\gamma^{-1}$. Dashed curve is the pure volume contribution (\ref{dIvol-amorph}), (\ref{Phi-def}). Dot-dashed, volume two plus single-boundary contributions (\ref{1b-amorph-def}), (\ref{B-Goldman-def}). Solid curve is the sum of all contributions.
}
\end{figure}

In contrast to (\ref{A1-magn})--(\ref{A2-magn}) and (\ref{A1-def})--(\ref{A2-def}), contribution (\ref{A-amorph-def}) due to the interference between the boundaries is not oscillatory. That is natural, as long as those oscillations depend on the final electron deflection angle, but in the present problem it is random, and is averaged over, thereby erasing the oscillations. Therewith also disappears the need for the attenuation formfactor $F_{\perp}$, which may be put equal to unity.

The behavior of a typical highly-nondipole radiation spectrum evaluated by Eqs. (\ref{dIvol-amorph})--(\ref{A-amorph-def}), along with its partial contributions, is shown in Fig.~\ref{fig:Amorph-spectrum}. At $\omega\to0$, it tends to the value dictated by the infrared factorization theorem (see \cite{Bondarenco-Shulga} and refs. therein). In practice, at $\omega\to0$ there may also arise a bump due to transition radiation, which was not taken into account within the present treatment \cite{Klein}.

%Let us mention, however, that interference features manifest themselves also in the opposite, weakly nondipole radiation case. For radiation at electron passage through a finite slab of amorphous matter, that can be shown by carrying out an expansion in powers of $\gamma^2\overline{\chi^2}$ beyond the leading (dipole) order. To this end, one may employ the quadrupole formfactor for bremsstrahlung at double scattering defined by Eq. (84) of Ref. \cite{Bondarenco-Shulga}, and average over points of scattering in the amorphous target to get
%\begin{equation}
%\frac{dI}{d\omega}=\frac{2e^2}{3\pi}\gamma^2\overline{\chi^2}\left[1-\frac{3\gamma^2\overline{\chi^2}}{10}F_q\left(\frac{\omega T}{2\gamma^2}\right)+\mathcal{O}\left(\gamma^4\overline{\chi^4}\right)\right].
%\end{equation}
%Here
%\begin{equation}
%F_q(\Omega_T)=\frac{80}{\Omega_T^2}\int_0^{\infty}\frac{duu}{(1+u)^8}\sin^2\left[\frac{\Omega_T}{2}(1+u)\right]
%\end{equation}
%(normalized by $F_q(0)=1$) is the quadrupole formfactor of radiation in the plate, describing suppression of the spectrum at $\frac{\omega T}{2\gamma^2}\lesssim 1$, in a way different from the Migdal formula pertinent to a highly nondipole regime.
%Here $\int_0^{\infty}d\Omega_T F_q(\Omega_T)\neq0$ (in fact, $F_q>0$), because averaging in a continuous medium permits coalescence of photon emission points: $t_1\to t_2$.

\section{Summary and outlook}

In spite of the diversity of possible electron scattering configurations
in finite targets, the corresponding radiation spectra admit similar decompositions, as described in Sec.~\ref{sec:decomposition}. Understanding of the underlying radiation mechanisms helps to determine which coherence length is relevant for which photon frequency range.

There are certain exceptions from the simplest variant of the decomposition theorem described here. Some of them were mentioned in Secs.~\ref{subsec:double-scattering} and \ref{sec:amorph-plate}. They merit more detailed investigation in the future.
%Let us add that when trajectories
%of the radiating electron are random, and the spectrum needs to be
%averaged over them, like for bremsstrahlung in a slab of amorphous
%matter, the final electron deflection angle, on which
%the period of oscillations of $A_1$ depends, is random, too.
%Averaging over it will erase the oscillations, and t

The relationship between the novel interference features discussed in the present article and jet and interjet, as well as time delay concepts of high-energy electrodynamics, permitting the neglect of the charged particle mass, opens new vistas for their investigation, with possible extensions to nuclear and hadron physics (cf., e.g., \cite{Eisberg-Yennie-Wilkinson,Dokshitzer}). The parallelism with radio- and antenna physics may bring new tools for the theory development. Investigations of radiation spectra behavior at low $\omega$ is also important from the applied point of view.

% As electron accelerator energies grow higher, there emerge more examples of this kind. The general prerequisite is some irregularity in electron motion -- sudden change of velocity vector or acceleration, and non-dipole ... -- ... .

%\begin{itemize}
%\item Jet effects can be essential even in angle-integral radiation spectra.
%\item Transition radiation exists for all kinds of boundaries, not necessarily between vacuum and atomic matter.
%\item There is `antenna' radiation from an ultra-relativistic, significantly deflecting electron.
%\item For separation of volume, edge and antenna contributions there exists a rigorous non-dipole decomposition.
%\item Weakly non-dipole radiation spectrum can be described by a quadrupole form factor.
%\end{itemize}

\section*{Acknowledgments}

This work was supported in part by the National Academy of Sciences
of Ukraine (Project CO-1-8/2017)  and the Ministry of Education and
Science of Ukraine (Project No. 0115U000473).

\end{document}